\documentclass[reqno,12pt,a4paper]{amsart}

\usepackage{mathrsfs}
\usepackage{amssymb}
\usepackage{amsfonts}
\usepackage{latexsym}
\usepackage{amsthm}
\usepackage{xcolor}
\usepackage{wrapfig}
\usepackage{floatflt}

\usepackage{graphicx}
\def\lb{\label}

\newcommand{\er}[1]{\textrm{(\ref{#1})}}

\begin{document}

%%%%%%%%%% Some definitions %%%%%%%%%%

%%%%%%%% Equations, theorems %%%%%%%%%
\renewcommand{\theequation}{\arabic{section}.\arabic{equation}}
\theoremstyle{plain}
\newtheorem{theorem}{\bf Theorem}[section]
\newtheorem{lemma}[theorem]{\bf Lemma}
\newtheorem{corollary}[theorem]{\bf Corollary}
\newtheorem{proposition}[theorem]{\bf Proposition}
\newtheorem{definition}[theorem]{\bf Definition}
\newtheorem{remark}[theorem]{\it Remark}
%\theoremstyle{remark}
%\newtheorem{remark}[theorem]{\bf Remark}

%%%%% Alphabet %%%%%
\def\a{\alpha}  \def\cA{{\mathcal A}}     \def\bA{{\bf A}}  \def\mA{{\mathscr A}}
\def\b{\beta}   \def\cB{{\mathcal B}}     \def\bB{{\bf B}}  \def\mB{{\mathscr B}}
\def\g{\gamma}  \def\cC{{\mathcal C}}     \def\bC{{\bf C}}  \def\mC{{\mathscr C}}
\def\G{\Gamma}  \def\cD{{\mathcal D}}     \def\bD{{\bf D}}  \def\mD{{\mathscr D}}
\def\d{\delta}  \def\cE{{\mathcal E}}     \def\bE{{\bf E}}  \def\mE{{\mathscr E}}
\def\D{\Delta}  \def\cF{{\mathcal F}}     \def\bF{{\bf F}}  \def\mF{{\mathscr F}}
\def\c{\chi}    \def\cG{{\mathcal G}}     \def\bG{{\bf G}}  \def\mG{{\mathscr G}}
\def\z{\zeta}   \def\cH{{\mathcal H}}     \def\bH{{\bf H}}  \def\mH{{\mathscr H}}
\def\e{\eta}    \def\cI{{\mathcal I}}     \def\bI{{\bf I}}  \def\mI{{\mathscr I}}
\def\p{\psi}    \def\cJ{{\mathcal J}}     \def\bJ{{\bf J}}  \def\mJ{{\mathscr J}}
\def\vT{\Theta} \def\cK{{\mathcal K}}     \def\bK{{\bf K}}  \def\mK{{\mathscr K}}
\def\k{\kappa}  \def\cL{{\mathcal L}}     \def\bL{{\bf L}}  \def\mL{{\mathscr L}}
\def\l{\lambda} \def\cM{{\mathcal M}}     \def\bM{{\bf M}}  \def\mM{{\mathscr M}}
\def\L{\Lambda} \def\cN{{\mathcal N}}     \def\bN{{\bf N}}  \def\mN{{\mathscr N}}
\def\m{\mu}     \def\cO{{\mathcal O}}     \def\bO{{\bf O}}  \def\mO{{\mathscr O}}
\def\n{\nu}     \def\cP{{\mathcal P}}     \def\bP{{\bf P}}  \def\mP{{\mathscr P}}
\def\r{\rho}    \def\cQ{{\mathcal Q}}     \def\bQ{{\bf Q}}  \def\mQ{{\mathscr Q}}
\def\s{\sigma}  \def\cR{{\mathcal R}}     \def\bR{{\bf R}}  \def\mR{{\mathscr R}}
\def\S{\Sigma}  \def\cS{{\mathcal S}}     \def\bS{{\bf S}}  \def\mS{{\mathscr S}}
\def\t{\tau}    \def\cT{{\mathcal T}}     \def\bT{{\bf T}}  \def\mT{{\mathscr T}}
\def\f{\phi}    \def\cU{{\mathcal U}}     \def\bU{{\bf U}}  \def\mU{{\mathscr U}}
\def\F{\Phi}    \def\cV{{\mathcal V}}     \def\bV{{\bf V}}  \def\mV{{\mathscr V}}
\def\P{\Psi}    \def\cW{{\mathcal W}}     \def\bW{{\bf W}}  \def\mW{{\mathscr W}}
\def\o{\omega}  \def\cX{{\mathcal X}}     \def\bX{{\bf X}}  \def\mX{{\mathscr X}}
\def\x{\xi}     \def\cY{{\mathcal Y}}     \def\bY{{\bf Y}}  \def\mY{{\mathscr Y}}
\def\X{\Xi}     \def\cZ{{\mathcal Z}}     \def\bZ{{\bf Z}}  \def\mZ{{\mathscr Z}}
\def\O{\Omega}

\newcommand{\gA}{\mathfrak{A}}
\newcommand{\gB}{\mathfrak{B}}
\newcommand{\gC}{\mathfrak{C}}
\newcommand{\gD}{\mathfrak{D}}
\newcommand{\gE}{\mathfrak{E}}
\newcommand{\gF}{\mathfrak{F}}
\newcommand{\gG}{\mathfrak{G}}
\newcommand{\gH}{\mathfrak{H}}
\newcommand{\gI}{\mathfrak{I}}
\newcommand{\gJ}{\mathfrak{J}}
\newcommand{\gK}{\mathfrak{K}}
\newcommand{\gL}{\mathfrak{L}}
\newcommand{\gM}{\mathfrak{M}}
\newcommand{\gN}{\mathfrak{N}}
\newcommand{\gO}{\mathfrak{O}}
\newcommand{\gP}{\mathfrak{P}}
\newcommand{\gQ}{\mathfrak{Q}}
\newcommand{\gR}{\mathfrak{R}}
\newcommand{\gS}{\mathfrak{S}}
\newcommand{\gT}{\mathfrak{T}}
\newcommand{\gU}{\mathfrak{U}}
\newcommand{\gV}{\mathfrak{V}}
\newcommand{\gW}{\mathfrak{W}}
\newcommand{\gX}{\mathfrak{X}}
\newcommand{\gY}{\mathfrak{Y}}
\newcommand{\gZ}{\mathfrak{Z}}

\newcommand{\ba}{\mbox{\boldmath$\alpha$}}
\newcommand{\bk}{\mbox{\boldmath$\kappa$}}
\newcommand{\bm}{\mbox{\boldmath$\mu$}}
\newcommand{\bet}{\mbox{\boldmath$\eta$}}

\def\ve{\varepsilon}   \def\vt{\vartheta}    \def\vp{\varphi}    \def\vk{\varkappa}

\def\Z{{\mathbb Z}}    \def\R{{\mathbb R}}   \def\C{{\mathbb C}}    \def\K{{\mathbb K}}
\def\T{{\mathbb T}}    \def\N{{\mathbb N}}   \def\dD{{\mathbb D}}

%%%%% Arrows %%%%%

\def\la{\leftarrow}              \def\ra{\rightarrow}            \def\Ra{\Rightarrow}
\def\ua{\uparrow}                \def\da{\downarrow}
\def\lra{\leftrightarrow}        \def\Lra{\Leftrightarrow}

%%%%% Typography %%%%%

\def\lt{\biggl}                  \def\rt{\biggr}
\def\ol{\overline}               \def\wt{\widetilde}
\def\no{\noindent}

%%%%% Math signs %%%%%

\let\ge\geqslant                 \let\le\leqslant
\def\lan{\langle}                \def\ran{\rangle}
\def\/{\over}                    \def\iy{\infty}
\def\sm{\setminus}               \def\es{\emptyset}
\def\ss{\subset}                 \def\ts{\times}
\def\pa{\partial}                \def\os{\oplus}
\def\om{\ominus}                 \def\ev{\equiv}
\def\iint{\int\!\!\!\int}        \def\iintt{\mathop{\int\!\!\int\!\!\dots\!\!\int}\limits}
\def\el2{\ell^{\,2}}             \def\1{1\!\!1}
\def\sh{\sharp}
\def\wh{\widehat}
\def\bs{\backslash}
%%%%% Math operations %%%%%

\def\all{\mathop{\mathrm{all}}\nolimits}
\def\Area{\mathop{\mathrm{Area}}\nolimits}
\def\arg{\mathop{\mathrm{arg}}\nolimits}
\def\const{\mathop{\mathrm{const}}\nolimits}
\def\det{\mathop{\mathrm{det}}\nolimits}
\def\diag{\mathop{\mathrm{diag}}\nolimits}
\def\diam{\mathop{\mathrm{diam}}\nolimits}
\def\dim{\mathop{\mathrm{dim}}\nolimits}
\def\dist{\mathop{\mathrm{dist}}\nolimits}
\def\Im{\mathop{\mathrm{Im}}\nolimits}
\def\Iso{\mathop{\mathrm{Iso}}\nolimits}
\def\Ker{\mathop{\mathrm{Ker}}\nolimits}
\def\Lip{\mathop{\mathrm{Lip}}\nolimits}
\def\rank{\mathop{\mathrm{rank}}\limits}
\def\Ran{\mathop{\mathrm{Ran}}\nolimits}
\def\Re{\mathop{\mathrm{Re}}\nolimits}
\def\Res{\mathop{\mathrm{Res}}\nolimits}
\def\res{\mathop{\mathrm{res}}\limits}
\def\sign{\mathop{\mathrm{sign}}\nolimits}
\def\span{\mathop{\mathrm{span}}\nolimits}
\def\supp{\mathop{\mathrm{supp}}\nolimits}
\def\Tr{\mathop{\mathrm{Tr}}\nolimits}
\def\BBox{\hspace{1mm}\vrule height6pt width5.5pt depth0pt \hspace{6pt}}
\def\where{\mathop{\mathrm{where}}\nolimits}
\def\as{\mathop{\mathrm{as}}\nolimits}

%%%%%%%%%%%%% specialities %%%%%%%%%%%%%%

\newcommand\nh[2]{\widehat{#1}\vphantom{#1}^{(#2)}}
%{{\mathop{#1}\limits^\wedge}\vphantom{#1}^{(#2)}}
\def\dia{\diamond}

\def\Oplus{\bigoplus\nolimits}

%%%%%%%%%%% End of definitions %%%%%%%%%%

%%%%% OLD OLD OLD

\def\qqq{\qquad}
\def\qq{\quad}
\let\ge\geqslant
\let\le\leqslant
\let\geq\geqslant
\let\leq\leqslant
\newcommand{\ca}{\begin{cases}}
\newcommand{\ac}{\end{cases}}
\newcommand{\ma}{\begin{pmatrix}}
\newcommand{\am}{\end{pmatrix}}
\renewcommand{\[}{\begin{equation}}
\renewcommand{\]}{\end{equation}}
\def\eq{\begin{equation}}
\def\qe{\end{equation}}
\def\[{\begin{equation}}
\def\bu{\bullet}

\title[{Wave propagation in periodic lattices with defects of smaller dimension}]
        {Wave propagation in periodic lattices with defects of smaller dimension}
\date{\today}

\def\Wr{\mathop{\rm Wr}\nolimits}
\def\BBox{\hspace{1mm}\vrule height6pt width5.5pt depth0pt \hspace{6pt}}

\def\Diag{\mathop{\rm Diag}\nolimits}

\date{\today}
\author[Anton Kutsenko]{Anton Kutsenko}
\address{Laboratoire de M\'ecanique Physique, UMR CNRS 5469,
Universit\'e Bordeaux 1, Talence 33405, France,  \qqq email \
aak@nxt.ru }

\begin{abstract}
The procedure of evaluating of the spectrum for discrete periodic
operators perturbed by operators of smaller dimensions is obtained.
This result allows to obtain propagative, guided, localised spectra
for different kind of physical operators on graphs with defects.
\end{abstract}

\maketitle

\section{Introduction}
\setcounter{equation}{0}

There are a lot of papers devoted to discrete periodic operators on
graphs, see reference in \cite{BKS}. At the same time there is the
strong interest on periodic structures with different kinds of
defects, see \cite{M}, \cite{OA}, \cite{CNJMM}, \cite{KS}, \cite{MS}
and \cite{MC}; about the continuous media see e.g. \cite{TMSD},
\cite{C}, \cite{ZGYZ} and \cite{PVDRDD}. Defects allows us to obtain
conductivity of the material for those frequencies (energies) at
which it was not in purely periodic structure. For example if we
take 2D homogeneous lattice then we have definite band-gap spectral
diagram of propagative waves. If we add some line defect we can
obtain new waves which propagate only along with this linear defect
and decay in perpendicular directions. The main goal of this paper
is to provide the general procedure of obtaining spectra appearing
after addition of the defects of smaller dimensions than the
original periodic lattice. Note that the technique described below
was used by the author (see \cite{K1}) to analyse the spectrum of
discrete wave equation in $2D$ lattice with linear defect and with
single inclusion to obtain propagative, guided and localised
spectra. The difference between this work and \cite{K1} is that in
\cite{K1} we provided detailed analysis of special kind of lattices
with defects, here we prove the general theorem and provide the
overall procedure to find the defect modes (without a detailed study
of its behavior) for a wide class of periodic lattices with
inclusions.

The work is organized as follows: Section \ref{S2} contains known
basic definitions and results from the theory of periodic operators
on graphs. Section \ref{S3} contains the procedure of obtaining the
spectrum of periodic operators perturbed by operators of smaller
dimensions.

Now, with help of definitions and results from Sections \ref{S2},
\ref{S3}, we present briefly the main results of the current work.
Consider $N$-periodic lattice $\pmb{\G}$ and $N$-periodic family of
operators $\cA(\o)$ acting on $\ell^2(\pmb{\G})$ with the parameter
$\o$. Physically $\o$ can be frequency, energy and so on, the
dependence on $\o$ is usually smooth or analytic. One of the main
problem is to analyse the set
\[\lb{i001}
 \O=\{\o:\ 0\in{\rm spec}(\cA(\o))\},
\]
which can be called as the set of possible states. The efficient
method for the study the spectrum of $N$-periodic operators is to
apply Bloch-Floquet transformation
$\hat\cA(\o):=\cF\cA(\o)\cF^{-1}$. The operators $\hat\cA(\o)$ acts
on $L^2_M([-\pi,\pi]^N)$ and are operators of multiplication by the
matrix-function
\[\lb{i002}
 \hat\cA(\o)\hat{\bf f}({\bf k})={\bf A}(\o,{\bf k})\hat{\bf f}({\bf
 k}),\ \ {\bf f}({\bf k})\in L^2_M
\]
with $M\ts M$ matrix ${\bf A}$ and ${\bf
k}=(k_1,..,k_N)\in[-\pi,\pi]^N$, here $M$ is a number of the nodes
in the unite cell of the graph. If ${\bf A}(\o,{\bf k})$ is a
continuous function then the set $\O$ can be described as follows
\[\lb{i003}
 \O=\{\o:\ \det{\bf A}(\o,{\bf k})=0\ {\rm for\ some}\ {\bf k}\}.
\]
The equation
\[\lb{i004}
 \det{\bf A}(\o,{\bf k})=0
\]
is called the dispersion equation, this defines the Bloch-Floquet
dispersion curves
\[\lb{i005}
 \o=\o({\bf k}),\ \ \ {\bf k}\in[-\pi,\pi]^N,
\]
the number of these curves is usually $M$ but can be larger if the
dependence on $\o$ of ${\bf A}$ is complicated. We have
$\O=\o([-\pi,\pi]^N)$, each point of the curves (branches) $\o({\bf
k})$ belongs to $\O$ and each point of $\O$ belongs at least to one
of the dispersion curve $\o({\bf k})$.

This is regarding to periodic operators, but what happens if we add
some periodic perturbations of $\cA$ of smaller dimension than $N$?
Such perturbations leads to study the family of the form
\[\lb{i006}
 \hat\cA(\o)\hat{\bf f}={\bf A}(\o,{\bf k})\hat{\bf f}+{\bf A}_1(\o,{\bf k})\langle\hat{\bf
 f}\rangle_1+...+{\bf A}_N(\o,{\bf k})\langle\hat{\bf
 f}\rangle_{1...N},
\]
where we denote
\[\lb{i007}
 \langle\hat{\bf f}\rangle_{i_1...i_j}=\frac1{(2\pi)^{\frac j2}}\int\limits_{[-\pi,\pi]^j}\hat{\bf
 f}dk_{i_1}...dk_{i_j}.
\]
The spectrum of operators $\hat\cA(\o)$ \er{i006} is well described
in the Theorem \ref{T1} below. Using this result we can determine
the Bloch-Floquet dispersion curves of smaller dimension than
\er{i005}. The procedure of finding the dispersion relations
consists of $N+1$ steps.

{\it Step 1.} The matrix
\[\lb{i008}
 {\bf B}_0(\o,{\bf k}):={\bf A}(\o,{\bf
k})
\]
defines the dispersion curves of the dimension $N$
\[\lb{i009}
 \det{\bf B}_0(\o,{\bf k})=0\ \ \Rightarrow\ \ \o=\o_0({\bf k}),\ \
 {\bf k}=(k_1,..,k_N)\in[-\pi,\pi]^N.
\]
(Note that $\o_0$ coincides with \er{i005})

{\it Step 2.} The matrix
\[\lb{i010}
 {\bf B}_1:={\bf I}+\langle{\bf B}^{-1}_0{\bf A}_1\rangle_1
\]
defines the dispersion curves of the dimension $N-1$
\[\lb{i011}
 \det{\bf B}_1(\o,{\bf k}_1)=0\ \ \Rightarrow\ \ \o=\o_1({\bf k}_1),\ \
 {\bf k}_1=(k_2,..,k_N)\in[-\pi,\pi]^{N-1}.
\]
The equation \er{i011} is not valid in the sets
\[\lb{i012}
 I_1({\bf k}_1)=\o_0([-\pi,\pi],k_2,..,k_N)
\]
because ${\bf B}_0^{-1}$ does not exists here. The sets ${\bf I}_1$
are the projections of the Bloch-Floquet dispersion (spectral)
curves on the plane $(\o,{\bf k}_1)$.

Continuing the process $N$ times we come to the last step

{\it Step N+1.} The matrix
\[\lb{i013}
 {\bf B}_N:={\bf I}+\langle{\bf B}^{-1}_{N-1}...{\bf B}^{-1}_0{\bf A}_N\rangle_{1...N}
\]
defines the dispersion curves of the dimension $0$ (these are
points, i.e. discrete spectrum)
\[\lb{i014}
 \det{\bf B}_N(\o)=0\ \ \Rightarrow\ \ \o_N.
\]
The equation \er{i011} is not valid in the sets
\[\lb{i015}
 I_N=I_{N-1}([-\pi,\pi])\cup\o_{N-1}([-\pi,\pi]),
\]
which are the projections of the Bloch-Floquet dispersion (spectral)
curves of the dimension greater or equal than $1$ on the axis $\o$.

The set $\O$ \er{i001} is union of all Bloch-Floquet branches
\[\lb{i016}
 \O=\bigcup_{j=0}^N\o_j([-\pi,\pi]^{N-j}).
\]
Note that the restriction to consider the dispersion curves
$\o_j({\bf k}_j)$ outside the projection $I_j({\bf k}_j)$ is
physically natural because the sets $I_j$ already consist of
spectral points $\o$ of higher dimension.

\section{Periodic lattices} \lb{S2}
\setcounter{equation}{0}

{\bf Def. 1. Periodic lattice.} We call the set $\pmb{\G}$ is the
$N$-periodic lattice with $M$-point unit cell if
\[\lb{001}
 \pmb{\G}=\bigcup_{{\bf n}\in\Z^N}\cL_{\bf n}
\]
where components $\cL_{\bf n}$ are disjoint
\[\lb{002}
 \cL_{{\bf n}_1}\cap\cL_{{\bf n}_2}=\es,\ \ {\bf n}_1\ne{\bf n}_2,
\]
and for each ${\bf n}$ there is the bijection
\[\lb{003}
 \vp_{\bf n}:\cL_{{\bf n}}\to[1,..,M].
\]

{\bf Def. 2. Group of translations.} For the lattice $\pmb{\G}$
\er{001} define the translation $\p_{\bf n}$ by
\[\lb{004}
 \p_{\bf n}:\pmb{\G}\to\pmb{\G},%\ \ {\bf n}\in\Z^N,
\]
\[\lb{005}
 \forall{\bf m}:\ \p_{\bf n}|_{\cL_{\bf m}}=\vp_{{\bf n}+{\bf
 m}}^{-1}\vp_{{\bf m}}.
\]
The translations satisfy
\[\lb{006}
 \p_{{\bf n}_1}\circ\p_{{\bf n}_2}=\p_{{\bf n}_1+{\bf n}_2}
\]
and the set of all translations
\[\lb{007}
 {\bf T}(\pmb{\G})=\{\p_{\bf n}:\ {\bf n}\in\Z^N\}
\]
is a group isomorphic to $\Z^{N}$.

{\bf Def. 3. Group of unitary translations.} For the lattice
$\pmb{\G}$ \er{001} define the unitary operators acting on the
Hilbert space of quadratic-summable functions $\ell^2(\pmb{\G})$
\[\lb{008}
 \cS_{\bf n}:\ell^2(\pmb{\G})\to\ell^2(\pmb{\G}),
\]
\[\lb{009}
 \cS_{{\bf n}}h=h\circ\p_{\bf n}.
\]
The set of all such operators
\[\lb{010}
 {\bf U}(\pmb{\G})=\{\cS_{\bf n}:\ {\bf n}\in\Z^N\}
\]
is a group isomorphic to $\Z^N$, since
\[\lb{011}
 \cS_{{\bf n}_1}\cS_{{\bf n}_2}=\cS_{{\bf n}_1+{\bf n}_2}.
\]

{\bf Def. 4. $N$-periodic operators.} The operator
\[\lb{012}
 \cA:\ell^2(\pmb{\G})\to\ell^2(\pmb{\G})
\]
is called $N$-periodic iff
\[\lb{013}
 \cA\cS=\cS\cA,\ \ \forall\cS\in{\bf U}(\pmb{\G}).
\]
{\it Remark.} It is sufficient to check the condition \er{013} only
for basis translations $\cS_{{\bf e}_j}$, $j=1,..,N$, where
\[\lb{014}
 {\bf e}_j=(\d_{ij})_{i=1}^N
\]
with Kronecker symbol $\d$.

{\bf Def. 5. Finite operators.} The operator
$\cA:\ell^2(\pmb{\G})\to\ell^2(\pmb{\G})$ is finite iff for any $h$
with finite support $\cA h$ has finite support too.

{\it Remark 1.} For $N$-periodic operator $\cA$ we need to check
finiteness only for the functions $\cA h_{m}$, $m=1,..,M$, where
$h_{m}$ is a function with the single support at the point
$\vp_{{\bf 0}}^{-1}(m)$.

{\it Remark 2.} Finite $N$-periodic operator is always bounded.

{\bf Def. 6. Bloch-Floquet transformation.} Define the unitary
operator $\cF$
\[\lb{015}
 \cF:\ell^2(\pmb{\G})\to L^2_M:=\os_{m=1}^ML^2([-\pi,\pi]^N),
\]
\[\lb{016}
 \cF h=(\hat h_m({\bf k}))_{m=1}^M,
\]
\[\lb{017}
 \hat h_m({\bf k})=\frac1{(2\pi)^{\frac M2}}\sum_{{\bf n}\in\Z^N}h(\vp_{\bf
 n}^{-1}(m))e^{i{\bf n}\cdot{\bf k}},
\]
where ${\bf k}=(k_j)_1^N\in[-\pi,\pi]^N$ and $\cdot$ is a scalar
product.

Floquet-Bloch transformation allows us to study $N$-periodic
operators efficiently, because

\begin{proposition} \lb{P1}
i) For $N$-periodic bounded operator $\cA$ the operator
$\hat\cA:=\cF\cA\cF^{-1}$ is an operator of multiplication by the
matrix
\[\lb{018}
 \forall\hat{\bf f}\in L^2_M:\ \hat\cA\hat{\bf f}={\bf A}({\bf k})\hat{\bf f},
\]
where $M\ts M$ matrix-function ${\bf A}$ is defined as
\[\lb{019}
 {\bf A}({\bf k}):=(\hat\cA\hat{\bf e}_1...\hat\cA\hat{\bf e}_N)
\]
with constant functions $\hat{\bf e}_j({\bf k})={\bf e}_j$ \er{014}.

ii) If $\cA$ is a finite $N$-periodic operator then the matrix ${\bf
A}({\bf k})$ is a finite sum
\[\lb{020}
 {\bf A}({\bf k})=\sum_{{\bf n}}e^{i{\bf n}\cdot{\bf k}}{\bf A}^{({\bf n})}
\]
with constant matrices ${\bf A}^{({\bf n})}$, ${\bf n}\in\Z^{N}$.

\end{proposition}
{\it Proof.} i) It is not difficult to show that the operator
\[\lb{021}
 \hat\cS_{\bf n}:=\cF^{-1}\cS_{\bf n}\cF=e^{-i{\bf n}\cdot{\bf
 k}}\cdot
\]
is the operator of multiplication by $e^{-i{\bf n}\cdot{\bf k}}$. So
if $\cA$ is $N$-periodic then $\hat\cA$ commutes with any
$\hat\cS_{\bf n}$, i.e.
\[\lb{022}
 \hat\cA e^{-i{\bf n}\cdot{\bf k}}\hat{\bf f}=e^{-i{\bf n}\cdot{\bf
 k}}\hat\cA\hat{\bf f}
\]
for any $\hat{\bf f}\in L^2_M$. Using linearity of $\hat\cA$ we
deduce that
\[\lb{023}
 \hat\cA r({\bf k})\hat{\bf f}=r({\bf k})\hat\cA\hat{\bf f}
\]
for any ${\bf f}\in L^2_M$ and for any trigonometric polynomial
$r({\bf k})$. Then \er{023} is fulfilled for any function $r({\bf
k})\in L^2$, since $\hat\cA$ is bounded and trigonometric
polynomials are dense in $L^2$. The identity \er{023} yields
\er{018} and \er{019}. The statement ii) immediately follows from
\er{019} and {\it Remark 1} after {\bf Def. 6.} \BBox

\section{Periodic operators of smaller dimensions} \lb{S3}
\setcounter{equation}{0}

{\bf Def. 7. Sublattices of smaller dimensions.} For $N$-periodic
lattice $\pmb{\G}$ introduce the sublattices:
\[\lb{200}
 \pmb{\G}_1=\bigcup_{{\bf n}\in\wt\Z^{N-1}}\cL_{\bf n}
\]
corresponding to the hyperplane
\[\lb{201}
 \wt\Z^{N-1}=\{{\bf n}\in\Z^N:\ {\bf n}\cdot{\bf e}_1=0\}
\]
and so on
\[\lb{202}
 \pmb{\G}_j=\bigcup_{{\bf n}\in\wt\Z^{N-j}}\cL_{\bf n}
\]
corresponding to the hyperplane
\[\lb{203}
 \wt\Z^{N-j}=\{{\bf n}\in\wt\Z^{N-j+1}:\ {\bf n}\cdot{\bf e}_j=0\}.
\]
For $j=N$ we have $\pmb{\G}_N=\cL_{\bf 0}$ is a finite set, for
$1\le j<N$ the set $\pmb{\G}_j$ is a $(N-j)$-periodic lattice. Note
that
\[\lb{204}
 \pmb{\G}_N\ss\pmb{\G}_{N-1}\ss...\ss\pmb{\G}_1\ss\pmb{\G}.
\]

{\bf Def. 8. Projectors on the sublattice.} Define the natural
projectors
\[\lb{205}
 \cP_j:\ell^2(\pmb{\G})\to\ell^2(\pmb{\G}_j)\ss\ell^2(\pmb{\G}).
\]
{\it Remark 1.} It is not difficult to show that
\[\lb{206}
 \hat\cP_j:=\cF\cP_j\cF^{-1}
\]
acts on any $\hat{\bf f}\in L^2_M$ as
\[\lb{207}
 \hat\cP_j\hat{\bf f}=\langle\hat{\bf f}\rangle_{1...j},
\]
where we denote
\[\lb{208}
 \langle\hat{\bf f}\rangle_{i_1...i_j}=\frac1{(2\pi)^{\frac j2}}\int\limits_{[-\pi,\pi]^j}\hat{\bf
 f}dk_{i_1}...dk_{i_j}.
\]

{\bf Def. 9. Periodic operators on the sublattice.} The operator
$\cA_j:\ell^2(\pmb{\G})\to\ell^2(\pmb{\G})$ is called
$(N-j)$-periodic on the sublattice $\pmb{\G}_j$ iff it acts on the
sublattice
\[\lb{209}
 \cP_j\cA_j\cP_j=\cA_j
\]
and $\cP_j\cA_j\cP_j$ restricted to the subspace
$\ell^2(\pmb{\G}_j)$ is $(N-j)$-periodic operator (see {\bf Def.4}).

Using properties of $\cP_j$ \er{205}-\er{207} it is not difficult to
show the analog of the Proposition \ref{P1} for the operators on
sublattices

\begin{proposition} \lb{P2}
i) For $(N-j)$-periodic bounded operator $\cA_j$ (see {\bf Def. 9})
the operator $\hat\cA_j:=\cF\cA_j\cF^{-1}$ is an operator of
multiplication by the matrix
\[\lb{210}
 \forall\hat{\bf f}\in L^2_M:\ \hat\cA_j\hat{\bf f}={\bf A}_j({\bf k})\langle\hat{\bf f}\rangle_{1...j},
\]
where $M\ts M$ matrix-function ${\bf A}_j$ is defined as
\[\lb{211}
 {\bf A}_j({\bf k}):=(\hat\cA_j\hat{\bf e}_1...\hat\cA_j\hat{\bf e}_N)
\]
with constant functions $\hat{\bf e}_i({\bf k})={\bf e}_i$ \er{014}.

ii) If, in addition, $\cA_j$ is a finite operator then the matrix
${\bf A}_j({\bf k})$ is a finite sum
\[\lb{212}
 {\bf A}_j({\bf k})=\sum_{{\bf n}}e^{i{\bf n}\cdot{\bf k}}{\bf A}_j^{({\bf n})}
\]
with constant matrices ${\bf A}_j^{({\bf n})}$ and ${\bf
n}\in\Z^{N-j}$.
\end{proposition}

Now we will study our main object: the spectrum of the $N$-periodic
operator $\cA$ on the lattice $\pmb{\G}$ perturbed by the
$(N-j)$-periodic operators $\cA_j$ on the sublattices $\pmb{\G}_j$.
Thus consider the operator
\[\lb{213}
 \cC=\cA+\cA_1+...+\cA_N.
\]
The spectrum of $\cC$ is the same as the spectrum of
$\hat\cC=\cF\cC\cF^{-1}$
\[\lb{214}
 \hat\cC=\hat\cA+\hat\cA_1+...+\hat\cA_N.
\]
Due to the Propositions \ref{P1},\ref{P2} the operator
$\hat\cC:L^2_M\to L^2_M$ has the following form:
\[\lb{215}
 \hat\cC\hat{\bf f}={\bf A}\hat{\bf f}+{\bf A}_1\langle\hat{\bf
 f}\rangle_1+...+{\bf A}_N\langle\hat{\bf f}\rangle_{1...N}
\]
for any $\hat{\bf f}\in L^2_M$. The matrices ${\bf A}_j$ depends on
${\bf k}=(k_1,..,k_N)$, but precisely does not depend on
$k_1,..,k_j$. The following Theorem provide the procedure of
verification $\l\in{\rm spec}(\hat\cC)$ or not.

\begin{theorem} \lb{T1}
Let $\hat\cC$ be defined in \er{215} with continuous
matrix-functions ${\bf A}_j$. For given $\l\in\C$ the condition
$\l\in{\rm spec}(\hat\cC)$ can be verified as follows: denote ${\bf
B}_0:={\bf A}-\l{\bf I}$ with identical matrix ${\bf I}$.

{\it Step 1.} If
\[\lb{216}
 \det{\bf B}_0({\bf k})=0\ \ {\rm for\ some}\ \ {\bf
 k}
\]
then $\l\in{\rm spec}(\hat\cC)$ else define the matrix
\[\lb{217}
 {\bf B}_1:={\bf I}+\langle{\bf B}_0^{-1}{\bf A}_1\rangle_1.
\]

{\it Step 2.} If
\[\lb{218}
 \det{\bf B}_1({\bf k})=0\ \ {\rm for\ some}\ \ {\bf
 k}
\]
then $\l\in{\rm spec}(\hat\cC)$ else define the matrix
\[\lb{219}
 {\bf B}_2:={\bf I}+\langle{\bf B}_1^{-1}{\bf B}_0^{-1}{\bf A}_2\rangle_{12}.
\]

....................................

{\it Step N.} If
\[\lb{220}
 \det{\bf B}_{N-1}({\bf k})=0\ \ {\rm for\ some}\ \ {\bf
 k}
\]
then $\l\in{\rm spec}(\hat\cC)$ else define the matrix
\[\lb{221}
 {\bf B}_N:={\bf I}+\langle{\bf B}_{N-1}^{-1}...{\bf B}_1^{-1}{\bf B}_0^{-1}{\bf A}_N\rangle_{12...N}.
\]

{\it Step N+1.} If
\[\lb{222}
 \det{\bf B}_{N}=0
\]
then $\l\in{\rm spec}(\hat\cC)$ else $\l\not\in{\rm spec}(\hat\cC)$.
\end{theorem}
{\it Proof.} Denoting $\hat\cC_0:=\hat\cC-\l$ we can rewrite the
condition $\l\in{\rm spec}(\hat\cC)$ as $0\in{\rm spec}(\hat\cC_0)$.
At the same time the condition $0\in{\rm spec}(\hat\cC_0)$ is
equivalent to that there is no $\hat\cC_0^{-1}$ or by the Banach
theorem that $\hat\cC_0$ is not a bijection, since $\hat\cC_0$ is
bounded.

{\it Step 1.} Suppose that $\det{\bf B}_0({\bf k}_0)=0$ for some
${\bf k}_0$. This means that there exists the vector ${\bf f}_0$
with quadratic norm $\|{\bf f}_0\|=1$ and with
\[\lb{223}
 {\bf B}_0({\bf k}_0){\bf f}_0={\bf 0}.
\]
We take sufficiently small $\d>0$ and take the function
\[\lb{224}
 \hat{\bf f}_0({\bf k})=\ca\frac1{\d^{N/2}}{\bf f}_0, & {\bf k}-\wt{\bf k}_0\in[0,\d]^N\\
                           {\bf 0}, & {\rm otherwise},
                        \ac
\]
where $\wt{\bf k}_0$ is close enough to ${\bf k}_0$ with the
condition $\wt{\bf k}_0+[0,\d]^N\ss[-\pi,\pi]^N$. Note that if ${\bf
k}_0\in(-\pi,\pi)^N$ then we can take $\wt{\bf k}_0={\bf k}_0$. We
have that $L^2_{M}$-norm of the function $\hat{\bf f}_0$ and its
integrals are
\[\lb{225}
 \|\hat{\bf f}_0\|=1,\ \ \|\langle\hat{\bf
 f}_0\rangle_1\|=\d^{\frac12},\ ...,\ \|\langle\hat{\bf
 f}_0\rangle_{1...N}\|=\d^{\frac N2}.
\]
Using \er{223}, \er{225} with the definition \er{215} we obtain
\[\lb{226}
 \|\hat\cC_0\hat{\bf f}_0\|\le\max\limits_{{\bf k}-\wt{\bf
 k}_0\in[0,\d]^N}\|{\bf B}_0({\bf k})-{\bf B}_0({\bf
 k}_0)\|+\d^{\frac12}\max\limits_{1\le j,\ {\bf k}}\|{\bf A}_j({\bf
 k})\|=:\ve(\d).
\]
The continuity of ${\bf B}_0({\bf k})$, ${\bf A}_j({\bf k})$ leads
to $\ve(\d)\to0$ for $\d\to0$. This means that $0\in{\rm
spec}(\hat\cC_0)$ because we found $\hat{\bf f}_0$ with $\|\hat{\bf
f}_0\|=1$ and with arbitrary small norm of $\hat\cC_0\hat{\bf f}_0$.

Now suppose that $\det{\bf B}_0({\bf k})\ne0$ for all ${\bf k}$.
Consider the equation
\[\lb{227}
 \hat\cC_0\hat{\bf f}={\bf B}_0\hat{\bf f}+{\bf A}_1\langle\hat{\bf
 f}\rangle_1+...+{\bf A}_N\langle\hat{\bf
 f}\rangle_{1...N}=\hat{\bf g}
\]
with some $\hat{\bf g}$. After multiplying \er{227} on ${\bf
B}_0^{-1}$
\[\lb{228}
 \hat{\bf f}+{\bf B}_0^{-1}{\bf A}_1\langle\hat{\bf
 f}\rangle_1+...+{\bf B}_0^{-1}{\bf A}_N\langle\hat{\bf
 f}\rangle_{1...N}={\bf B}_0^{-1}\hat{\bf g}
\]
and taking $\langle\cdot\rangle_1$ we obtain
\[\lb{229}
 \hat\cC_1\langle\hat{\bf f}\rangle_1:={\bf B}_1\langle\hat{\bf
 f}\rangle_1+...+\langle{\bf B}_0^{-1}{\bf A}_N\rangle_1\langle\langle\hat{\bf
 f}\rangle_1\rangle_{2...N}=\langle{\bf B}_0^{-1}\hat{\bf
 g}\rangle_1.
\]
Note that the operator $\hat\cC_1$ acts on
$L^2_M([-\pi,\pi]^{N-1})$. If $\hat\cC_0$ is a bijection then
$\hat\cC_1$ is a bijection too. The inverse statement is true also,
because we can uniquely reconstruct $\hat{\bf f}$ from
$\langle\hat{\bf f}\rangle_1$ by using \er{228}. So we conclude that
\[
 0\in{\rm spec}(\hat\cC_0)\ \Leftrightarrow\ 0\in{\rm
 spec}(\hat\cC_1).
\]
Now applying the {\it Step 1} to the operator $\hat\cC_1$ we finish
the Proof by induction. \BBox

\end{document}